# Observations of the long-lasting activity of the distant comets 29P Schwassmann-Wachmann 1, C/2003 WT42 (LINEAR) and C/2002 VQ94 (LINEAR).

Oleksandra V. Ivanova*[a], Yury V. Skorov[b,c], Pavlo P. Korsun[a], Viktor L. Afanasiev[d] and Jürgen Blum[b]

[a] – Main Astronomical Observatory of NAS of Ukraine, Akademika Zabolotnoho 27, 03680 Kyiv, Ukraine

* Corresponding Author. E-mail address: sandra@mao.kiev.ua

[b] – University of Braunschweig. Institute for Geophysics and Extraterrestrial Physics

Mendelssohn-Str. 3, D-38106 Braunschweig, Germany

[c] – Max Planck Institute for Solar System Research, Max-Planck-Str. 2, D-37191,Katlenburg-Lindau, Germany

[d] – Special Astrophysical Observatory of the Russian AS, Nizhnij Arkhyz, 369167, Russia

**Proposed Running Head**: Observations of the long-lasting activity of the distant comets 29P Schwassmann-Wachmann 1, C/2003 WT42 (LINEAR) and C/2002 VQ94 (LINEAR).

**Editorial correspondence to:**

Dr. Oleksandra Ivanova

Main Astronomical Observatory of NAS of Ukraine

Akademika Zabolotnoho str. 27

03680 Kyiv

Ukraine

Phone: +380 44 526-47-69

E-Mail address: sandra@mao.kiev.ua

# ABSTRACT

We investigated three comets, which are active at large heliocentric distances, using observations obtained at the 6-m BTA telescope (SAO RAS, Russia) in the photometric mode of the focal reducer SCORPIO. The three comets, 29P/Schwassmann–Wachmann 1, C/2003 WT42 (LINEAR) and C/2002 VQ94 (LINEAR), were observed after their perihelion passages at heliocentric distances between 5.5 and 7.08 AU. The dust production rates in terms of $Af\rho$ was measured for these comets. Using these retrieved values, an average dust production rate was derived under different model assumptions. A tentative calculation of the total mass loss of the cometary nucleus within a certain observation period was executed. We calculated the corresponding thickness of the depleted uppermost layer where high-volatile ice completely sublimed. The results obtained in our study strongly support the idea that the observed activity of the comet SW1 needs a permanent demolition of the upper surface layers.

# 1. Introduction

Cometary nuclei consist of ices intermixed with dust grains and are thought to be the least modified solar system bodies remaining from the time of planetary formation. Therefore investigation of the cometary composition is important for understanding the processes that occurred during the formation of our solar system (Greenberg, 1982; Li and Greenberg, 1997). The automated all-sky searches for Near-Earth-Objects such as the LONEOS Survey (Howell et al., 1996) and the NEAT Survey (Pravdo et al., 1999), the LINEAR program (Stokes et al., 2000), the Catalina Sky Survey (Larson et al., 1998), and others discover new comets arriving in the inner Solar System regularly. Some of these comets show significant activity beyond the orbit of Jupiter.

Moreover, these comets demonstrate various patterns of activity (Lowry, 2003; Bauer et al., 2003; Meech, 2009). Rich molecular spectra above the underlining continuum are detected in two comets. The tails are not observed in these comets and their morphologies are defined by asymmetric comae with jets features (Stansberry et al., 2004; Korsun, 2006; Korsun, 2008; Trigo-Rodríguez et al., 2008). Other distant comets do not demonstrate gas emissions above the reflected solar continuum in the optical spectra. Moreover some of them have long dust tails. (Roemer, 1962; Belton, 1965; Korsun et al., 2003; Meech, 2009).

The activity of the comets at large heliocentric distances cannot be explained in the frame of the standard model (Whipple, 1972), where the sublimation of water ice heated by solar radiation is the primary cause of the activity of the cometary nucleus. The problem which physical mechanisms explain the distant cometary activity was debated in a number of publications over the past decade and several possible mechanisms were proposed to explain the activity of distant comets (see, e.g., review by Gronkowski, 2005). The most popular presumable sources of energy are: i) the sublimation of more volatile species than $H_2O$ such as CO and/or $CO_2$ ice (e.g. Hopis and Mendis, 1981; Prialnik, Bar-Num, 1992, Hughes, 1992), ii) polymerization of HCN (Rettig et. al., 1992), iii) crystallization of the amorphous water ice (Gronkowski et al., 1998; de Sanctis et al., 2002; Prialnik, 1992) and iv) annealing of the amorphous water ice (Meech et al., 2009). However, there is no generally accepted mechanism to interpret the activity of the observed distant comets, and therefore, further long-term observations of the distant comets are desirable to explain their activity. A program of photometric and spectral studies of the comets showing a high level activity beyond Jupiter's orbit was started at the 6-m telescope BTA (SAO RAS, Russia) in 2006. Both images and spectra of several distant comets were obtained during observations made from 2006 to 2009. Here we present the photometric observations of three distant comets and the corresponding their analysis.

The Centaur 29P/Schwassmann-Wachmann 1 exhibits a considerable level of activity (so-called outbursts) since its discovery in 1925. The comet moves along a near-circular orbit. The orbital period of the comet is equal to 14.9 years (Herget, 1961). Trigo-Rodriguez at al. (2008) detected 28 outbursts during their 2002−2007 observational campaign. Typical 29P/Schwassmann-Wachmann 1 outbursts occur with a frequency of 7.3 outbursts per year and are characterized by the sudden increase of 1 to 4 mag in brightness. There was no sign of periodicity in the detected outburst, however the reason for the non-periodic character of the outbursts has not been found yet.

Object A/2002 VQ94 (LINEAR) was discovered as an asteroid by the LINEAR team on November 11.24, 2002 at a heliocentric distance of 10.02 AU (Marsden, 2002). Observations, were made in the end of August 2003, when the heliocentric distance of the comet was 8.9 AU, revealed a prominent coma with a fanlike morphology (Green, 2003). Thus, the object was put on a cometary list as C/2002 VQ94 (LINEAR) (Parker, 2003). A more extended asymmetric coma was observed through V and R filters at heliocentric distances of 6.8 AU and 7.3 AU, respectively (Korsun et al., 2006, 2008). Both comets, C/2002 VQ94 (LINEAR) (further VQ94) and 29P/Schwassmann–Wachmann-1 (further SW1), demonstrate similar morphologies. The spectra of the comets are emission rich. For the first time the $CO^+$ lines were identified during observations of the comet SW1 carried out in 1978 and 1979 by Cochran et al. (1980, 1982). Further, Jockers et al. (1992) obtained images of SW1 in the $CO^+$ line and investigated the morphology of the ionic component. Later, the $N_2^+$ emissions above the continuum were detected in the spectra of the comet SW1 in addition to the identified lines and $CO^+$, CN, $N_2^+$ and the $C_3$ emissions were discovered in the spectra of comet VQ94 (Korsun et al., 2006, 2008).

Object C/2003 WT42 (LINEAR) (further WT42) was first identified as an unusual asteroid. The 18th magnitude object was discovered by the LINEAR team on November 19.26, 2003 at a heliocentric distance of 8.1 AU (Marsden, 2004). Observations were made in mid January 2004 demonstrated the presence of a coma, confirming the object as a comet. The comet moves along a hyperbolic orbit with a perihelion distance of 5.19 AU. Marsden (2004), analyzing the barycentric values, suggested that this could be a "new" comet from the Oort cloud. The comet showed long-term activity since early 2005. Spectroscopic observations of the comet did not reveal any gas emissions in the spectra (Korsun et al., 2009, (submitted to the editor)). The digital processing of the images indicated that, besides isotropic outgassing, there was an outflow of cometary matter from two active areas localized on the nucleus (Korsun at al., 2009, (submitted to the editor)). The comet exhibited a long dust tail during the observation. Fig 1 shows the three cometary orbits related to the ecliptic plane.

[**Figure 1**]

The analysis of observation of cometary activities at large heliocentric distances requires the use of theoretical models describing the processes in the upper layers of the nucleus. A set of numerical models were proposed allowing computer investigations of the energy and mass transfer in porous cosmic bodies. (Skorov at al., 1995; Skorov at al., 2001; Davidsson et al., 2004). Hereafter we discuss some contingencies arising from such a computer analysis. Primarily, we focus our study on the explanation of the long-term activity of the distant comets listed above.

## 2. Observational techniques and image processing

In this paper we describe observations of three distant comets (SW1, VQ94 and WT42) between 2006 and 2009. Our observations were carried out at the 6-m telescope BTA (SAO RAS, Russia) during several observing runs. Photometric images of comets were obtained using the focal reducer SCORPIO attached to the prime focus of the telescope (Afanasiev and Moiseev, 2005). A CCD chip EEV-42-40 of 2048×2048 pixels was used as a detector. The full field of view of the CCD is 6.1′×6.1′. We applied a 2x2 binning during the observation therefore the image scale was 0.36″/pix. The three comets were observed through broadband *B, V* and *R* filters. There were two sets of observations for comet SW1. The first was in December 2006, when the heliocentric and geocentric distances of the comet were 5.865 and 4.947 AU, the second was in March 29, 2009, when the heliocentric and geocentric distances of the comet were 6.118 and 5.805 AU, respectively. Comet VQ94 was observed at heliocentric and geocentric distances equal to 7.084 and 7.332 AU in December 12, 2006. The images of comet WT42 were obtained in December 16, 2006 at heliocentric and geocentric distances equal to 5.519 and 5.598 AU, respectively. Despite the fact that the comets were observed at heliocentric distances at which the sublimation of the water ice is too low, the comets showed considerable activity. Details on the photometric observations are collected in Table 1. Fig. 2 is an illustration of the observed data. The series of homogeneous data were summed in a standard suitable form for further use. For this purpose a robust statistics was applied to the raw data. The reductions of the raw data, which included bias subtraction, flat field correction and cleaning from cosmic ray tracks, were thoroughly done. The bias was removed by subtracting an averaged frame with zero exposure time, the morning sky was exposed to provide flat field corrections for the non-uniform sensitivity of the CCD chip, and the cosmic ray tracks were removed when we computed the composite image from the individual ones. The night sky level of each individual frame was estimated from those parts of the frame, which were not covered by the cometary coma and tail. To provide the absolute photometric calibration we used the spectral atmospheric transparency at the Special Astronomical Observatory obtained by Kartasheva and Chunakova (1978) and our observations of the spectrophotometric standard stars BD+75d325, BD+33d325, G193-74, Feige 56 (Oke, 1990) and the Landolt photometric standard star PG1047+003 (Landolt, 1992).

[Figure 2]

[Table 1]

## 3. Results

### 3.1 Photometry: Methods of image analysis

The photometric observations obtained in the different spectral bands were used to estimate the dust production rate of the distant comets. The continuum flux was measured within circular apertures centered on the comet. Measurement data were converted into energety units via the equation (Busarev, 1992):

$$F_c(\lambda) = F_s(\lambda) \cdot \frac{I_c(\lambda)}{I_s(\lambda)} \cdot P^{-\Delta M}(\lambda), \quad (1)$$

where $F_s$ is the energy distribution in the spectra of the standard star convolved with the filter transmitting curves; $I_s$ and $I_c$ are the measured fluxes of the star and the comet in counts respectively; $P$ is the coefficient of the sky transparency; $\Delta M$ is the difference between the comet's and star's airmasses.

In order to compare the measurements of the dust productivity taken at different epochs, observing sites and with different viewing geometries, the product of the average grain albedo $A$ and the filling factor of grains $f$ is widely used (A'Hearn and Schleicher, 1984). We calculated $Af\rho$ using the equation:

$$Af\rho = \frac{4 \cdot (r/1AU)^2 \cdot \Delta^2}{\rho^2} \frac{F_c}{F_{sun}}, \quad (2)$$

where $A$ is average albedo of the particles, $f$ is the filling factor of the grains within an aperture $\rho$ projected on the comet distance and expressed in cm, $r$ is the heliocentric distance in AU, $\Delta$ is the geocentric distance in cm, $F_{sun}$ is the flux of the Sun at 1AU convolved with the filter transmitting curve (Neckel at al., 1984). The relation (2) is commonly applied when both, the dust production rate and the velocities of the ejected particles, are constants (a simple steady-state coma model). For such an ideal steady-state coma, $Af\rho$ is aperture independent and can be used to derive the lower limit of the dust production rate (Bauer et al., 2003) and to compare the activities of different comets. We calculate the $Af\rho$ values for a number of apertures from 5 to 20 pixels. The final error of the *Afρ* value is estimated as the square of the sum of the uncertainty of the cometary flux introduced above and the uncertainty of the solar energy flux (Arvesen, 1969). Then, a lower limit to the dust production rate can be calculated by

$$Q = Af\rho \cdot \frac{4/3 \cdot \pi \cdot a^3 \cdot v \cdot \sigma}{p \cdot a^2}, \quad (3)$$

where *a* is the grain radius, *p* is the grain albedo, *v* is the grain ejection velocity and *σ* is the grain density. We assume that for the comets under consideration the average optical albedo and the density of the grain material are fixed at *p*=0.04 and $\sigma \approx 1$ g·cm$^{-3}$ (Jewitt, 1990; Meech et al., 1993;). This albedo value is often used for the comet SW1. As to the density of particles in comets, Britt et al. (2006) showed that this value can vary in a wide range, depending on the composition of the dust particles, their porosity, size and so forth. In this paper we adopt a typical value of the density, which is often used for various estimations in comets.

Here, we use a simple model in which dust particles are described as solid spheres with radius from 1 μm to 10 μm and with dust size distribution of 2.5<a<3.0 (Jockers, 1997). The corresponding limits of ejection velocity are 50 m·s$^{-1}$ and 10 m·s$^{-1}$. The grain radii and velocities are chosen as a result of the quantitative analysis of the data reported for different comets by Fulle (1992), Fulle et al. (1998), Korsun et al. (2003, 2009). Obviously we make a gross simplification when we characterize the dust grains by a single grain radius, but this idealization seems reasonable due to inherent restrictions of the *Afρ* method. Because we plan to calculate the limits of the total dust production rate of the investigated distant comets, the well-founded estimation of the dust size limits seems appropriate. An additional argument can be found from the lifetime estimation of the grains. Mukai (1986) concluded that at a heliocentric distance of 7 AU the lifetime of dirty icy dust particles with a radius from 1 to 10 μm is few hours. Because both SW1 and VQ94 show no dust tails, one can assume that their atmospheres contain dirty icy particles of less than 10 μm in size. Otherwise a dust tail (as in the case of comet WT42) would be formed by larger particles with a sufficiently long lifetime. Sophisticated dynamical modeling (Sekanina, 1975; Sekanina, 1982; Meech, 1993; Korsun and Chorny, 2003) shows that such tails are formed by a continuous process of ejection of large grains with low ejection velocity and, hence, with long lifetime. In order to check how the retrieved results are sensitive to the discussed above simplification of the model, we tested additionaly the case in which the dust particles are treated as fluffy one-size fractal grains with a radius of 1 μm. It should be noted that in this case the equation for $Q$ loses its accuracy, because the fractal particle mass grows weaker with size than for a solid sphere. Due to a higher efficiency of dust-gas interaction and, thus, higher dust acceleration, the corresponding ejection velocity should be also much higher than for the case of solid grains. We use as a test value velocity of 500 m·s$^{-1}$.

The main results presented in this section were derived from the images obtained in December 2006 and March 2009, when the comets were located at post-perihelion distances. The calculated $log(Af\rho)$-$log(\rho)$ and $Q$-$\rho$ relations for the projected distances from about 15000 km to 57000 km from the nucleus are plotted in Figs. 3 and. 4(a,b).

**[Figure 3]**

**[Figure 4a, Figure 4b]**

To calculate the dust productivity, we suppose that particles are emitted isotropically from the nucleus and move radially outward with constant velocity. The spatial density of the dust grains, $N_d \propto 1/r^2$, and the observed brightness decreases as $1/r$, where $r$ is the distance from the cometary nucleus. In this case, the logarithmic surface brightness gradient would have the slope of -1 (Lowry and Fitzsimmons, 2005). The slightly steeper gradient (Fig. 3) than expected in the case of uniform outgassing, can be caused by several effects, such as the radiation pressure accelerating grains while they move outwards, the sublimation of grains and their gradual fading, the possible existence of jets creating non-isotropic features disturbing the uniform brightness distribution (Meech et al., 1993; Jockers and Bonev, 1997; Lowry and Fitzsimmons, 2005). Quantitative discussion of these mechanisms is beyond the scope of this work. We note only that $Af\rho$ becomes aperture dependent in this case. However, one can get physically reasonable results even in the frame of our simple model.

It is significant that the objects under consideration evidently show considerable dust activity in the post-perihelion period. Average values of *Afρ* are 114.3 ± 1.2 cm for VQ94, 762.3 ± 2.4 cm for WT42, 1250.5 ± 3.2 cm for 29P in December 2006 and 988.3 ± 3.4 cm in March 2009. These *Afρ* values are typical for dynamically new comets (Meech et al., 2009; Epifani et al., 2009). Meech et al. (2009) analyzed the activities of three dynamically new distant comets (C/1992 J2, C/2001 G1, C/2003 A2) and concluded that these comets are visibly brighter and fade more slowly after their perihelion than most of the short period comets. Our comets WT42 and VQ94 demonstrate this slow post-perihelion fading too.

The monitoring of the comet SW1 (Trigo-Rodriguez et al., 2008) showed that the comet brightness decreased considerably at the end of 2006. We observed a high level of dust productivity of the comet SW1 in 2006 that might be explained by a high jets activity of this comet at the observational time. In 2009 we retrieved a lower value of the dust productivity. The results of SW1 observations from the CARA project data base (Cometary Archives for Amateur Astronomers, http://www.uai.it) in March 2009 demonstrated that the brightness of the comet was much higher than in 2006. The observed variations of *Afρ* may be connected with variations in the number of jets (5–6 jets were observed in 2006 (Korsun et al., 2008); whereas in 2007 only 2-3 jets (Ivanova et al., 2009) and 3 jets were observed in 2009 (Ivanova, private communication)). Thus we can suppose that the observed dust activity of SW1 correlates at least partially with the jet number, rather than with the general brightness of the comet.

The comet WT42 showed a very high level of the activity in the pre-perihelion and the post-perihelion periods. *Afρ* values of WT42 demonstrated high production rate amounting to 3500 cm when the comet was at a distance of 6.4 AU from the Sun (Korsun et al., submitted). Our results showed also that the comet had a lower value of the dust productivity in December 2006 than in March 2006, when the comet demonstrated high level activity. The results are in good agreement with those derived by other investigators (Cometary Archives for Amateur Astronomers, http://www.uai.it). Although the dust production of WT42 is higher than the production of comet VQ94 (Fig.3), both comets had similar rate of post perihelion fading.

Unfortunately we cannot compare our dust production rate (Fig. 4(a, b)) for comets VQ94 and WT42 with analogue observational data others observers. The observations of these comets were started only recently, and at the moment we have no extensive evidence for cometary activities and estimations of the dust production rates. The results of the dust mass loss rate for the comet 29P SW1 were obtained from analysis of the dust coma (Fulle, 1992; Jewitt, 1990; Stansberry et al., 2004; Moreno, 2009). Fulle (1992) derived a loss rate of $600 \pm 300$ kg s$^{-1}$, while Jewitt (1990) derived only 10 kg s$^{-1}$, Stansberry et al. (2004) obtained an upper limit of 50 kg s$^{-1}$, and Moreno (2009) derived $300 \pm 100$ kg s$^{-1}$. Our results are close to the estimates of Fulle (1992) and Moreno (2009).

# 4. Discussion

The activity of the comets considered here is not a unique or rarely observed phenomenon. On the contrary, the number of space objects showing a distant activity is steadily increasing (nowadays, more than 100 comets are under observations, while there are 2-3 new comets being discovered each year). Occasional outbursts (e.g., the activity of comet Hale-Bopp at the extra large heliocentric distance (at 6 AU) (Sekanina et al., 1996), (at 25. 7 AU) (Szabo et al., 2008)) as well as long-term activity, accompanied by the release of considerable amounts of dust and gas, were observed and are described in many papers (Biver et al., 1996; Sekanina et al., 1996; Rauer et al., 1997; Prialnik, 1997; Jewitt et al., 1999). We restrict our consideration to physical processes that may lead to the long-term high activity of cometary nuclei.

The above results of our observations not only demonstrate the high activity of the considered comets at large heliocentric distances. An important specific feature of this activity is its long-term nature, in contrast to the often observed outburst activity. This feature allows discussing in more detail the physical processes associated with this activity. The discussion of the distant cometary activity usually focuses on finding proper sources of energy required to release particles from the cometary nucleus that, in turn, leads to the formation of the observed coma. We also dwell on this issue below. However, it is the long-term activity of remote comets which drew our attention to the other side of the problem that has not been addressed in studies so far. By this we mean the study of the structural transformation of the uppermost layers of a cometary nucleus dealing with its activity. This side of the problem may be even more important for the long-time activity discussion, although it is not so obvious, and we start from that.

Using the obtained values of the dust production rate we can make a preliminary calculation of the total mass loss of the cometary nucleus within a certain observation period, assuming that the sublimation of a high-volatile ice is necessary to drive the dust particles away from the nucleus. Even a rough quantitative estimation allows us to make a tentative analysis of the possible scenarios of the long-term cometary activity at large heliocentric distances. To calculate the resulting mass loss one has to specify the cometary nucleus composition and its size. In addition to that, we need to take into account the total observation time of the comet. Comet SW1 has been observed for 82 years from its first discovery (Roemer, 1958) and displayed a high activity over many years (Cochran et al., 1980). We suppose that comet VQ94 had a visible activity during 6 years (Green, 2003), and comet WT42 was active during 5 years at least (Marsden, 2004). All these comets have nuclei larger than those of typical short-period comets. The nucleus radius of comet SW1 lies in the range of 15-44 km as obtained from visual photometry (Meech et al., 1993) and from the analysis of its thermal emission (Cruikshank and Brown, 1983; Stansberry et al., 2004). The radius of comet VQ94 computed by Jewitt (2005) is about 40 km (assuming an albedo of $p = 0.04$). There is no estimation of the nucleus radius for the comet WT42, and we assumed that the nucleus radius is about 30 km as an average value of the SW1 and VQ94 radii. At present the composition of cometary nuclei are determined insufficiently. As we know from observations coma composition may vary in a wide range from one comet to the other (Mumma et al., 2003; Bockelee-Morvan et al., 2004). However, we can say distinctly that all considered nuclei contain both ices and dust. Because of the composition uncertainty we estimate the total mass loss for several model variants: 1) a model with a high content of dust for which we assume a mass ratio of dust/gas of 2; and 2) a model with a low content of dust (mass ratio of dust/gas of 0.5). For each model we examine, in turn, two options: a) a nucleus containing a large fraction of high-volatile components (Fernandez and Jockers, 1983; Houpis et al., 1981), b) a nucleus containing a high-volatile fraction "typical" for the short period comets (Mumma et al., 2003). In order to derive the total mass loss, an idealized one-dimensional thermophysical model of cometary nucleus is used. The model nucleus has a spherical shape and a macroscopically homogeneous composition. The last assumption means that all physical characteristics of the model (such as density, heat capacity and so on) can be determined via standard expressions (Davidsson and Skorov, 2004). The "fast rotator" approximation (i.e. averaging the energy input during one full rotation of the cometary nucleus) is applied to specify the energy budget on the cometary surface. The final thickness of the layer depleted by high-volatile components is calculated from the conservation of mass. An activity depression is roughly estimated for the simple case, in which only high-volatile ice is sublimated and the porous uppermost layer is described as a bundle of cylindrical capillaries with scattering walls. A time evolution of such a layer is simulated by the Clausing formula. A more detailed description of the model and discussion of theoretical aspects of the considered problem will be presented in a forthcoming publication. We present our calculated values of the integral mass loss for the three considered comets in Table 2.

[Table 2]

Estimates of the total losses of dust presented the Table 2 clearly display that for all considered comets there is a finite depth of the uppermost layer devoid of ice and other high-volatile constituents. The thickness of this layer is at least a few centimeters (see Tab. 2). Taking into account that the size of the released particles in our model does not exceed 10 microns, we find that the thickness of the layer is several orders of magnitude greater than the characteristic size of the pores. This conclusion gives us a strong argument in support of the idea that the observed activity of the comet SW1 needs permanent demolition of the upper surface layers.

The same outcome can be obtained from the consideration of the migration of the sublimation products through a porous ice layer. It is easy to show that the temperature of the cometary surface at the considered distances is below 140K which is much lower than the temperature range in which water ice sublimes actively. It means that the products of the sublimation of more volatile species (e.g. CO, $CO_2$) migrate through a region where only pure water ice is conserved, as well as through a reflecting porous dust layer. As a result, one can trust that the general physical effects observed for the sublimation of water ice beneath the porous dust layer, as well as the mathematical approaches developed for the solution of this task are valid. Applying standard formulae describing a free molecular flow through porous media (e.g. the Clausing formula), we obtain that generally the layer of the passive water ice reduces the effective production rate inversely to its dimensionless thickness (where the last value is determined as the ratio of the layer thickness to its effective pore size). The characteristic time when the sublimation should vanish in such a model is several days only, that is much shorter than the observed time of high activity of comet SW1. The same analysis is completely applicable to the other comets considered here. We see that the thickness of the uppermost layer devoid of high-volatile ice components can vary from one model to other, but in all cases it stops very quickly and an effective ice sublimation from beneath ceases.

As we noted above, finding the proper sources of energy responsible for the observed activity of distant comets is of particular interest for cometary researchers. A detailed comparative analysis of the usually used hypotheses involved to explain a permanent activity of distant comets was recently published by Gronkowski (2007). The author considered different sources of energy required for the dust and gas release at large heliocentric distances, such as meteoritic impacts, crystallization of amorphous water ice and polymerization of hydrogen cyanide, and concluded that the water ice transformation appears to be the most promising one. Starting from Patashnick (1974) and Smoluchowski (1981) the idea that the amorphous-to-cubic phase transition of solid water ice is the most reasonable physical process responsible for the observed effects was used in numerous thermo-physical models of cometary nuclei and is still popular in modern cometary physics. We refer here to the following important publications and references therein (Prialnik 1992; Enzian et al. 1997). The key point for all cited papers is that a solid state phase transformation is treated as an exothermic reaction leading to an internal energy release. The other particular feature that makes this idea attractive for cometary researchers is the possible release of minor more volatile components (e.g. CO or $CO_2$) trapped in amorphous ice during this phase transformation (Bar-Nun et al. 1985; Prialnik and Bar-Nun 1990; Bar-Nun and Laufer 2003).

However, we can hardly rely on this idea completely. The first problem arises concerning admixtures of cometary material. The crystallization is indeed an exothermic transformation for pure water ice, but this situation may change dramatically if ice with dopants is considered. As experiments demonstrate that a net energy release during crystallization may be even negative for different physically reasonable cases (Sandford and Allamandola 1990, Kouchi and Sirono 2001). Thus, we support the approach suggested by Enzian et al. (1997), where the net energy release is treated as a model parameter depending on the material composition. The authors presented a sophisticated thermo-physical model of the SW1 nucleus including multidimensional heat transport, crystallization of amorphous water ice, diffusion and release of minor volatiles. They concluded that a total surface erosion is required in order to keep the amorphous ice close to the surface as well as to conserve a significant gas flux of carbon monoxide. We see that the usually suggested runaway crystallisation leads to quick and continuous lessening of the gas activity. It is interesting to note that Smoluchowski (1981) pointed out that crystallization of water ice might be important for SW1 "if layers of the crystallized ice peel off". This is the second problem arising when one applies the idea about a solid-state transformation of water ice to investigate the behavior of the comet SW1. This obstacle seems more serious, however: if in the former case we cannot evaluate accurately the net energy release and, hence, the efficiency of this mechanism, in the latter case we cannot even suggest candidates that could totally peel the surface layers off. Indeed, the most natural candidate is the pressure of sublimating high-volatile gases. To a crude approximation one can estimate the corresponding gas pressure from the expression for saturation pressure as a function of equilibrium temperature. In all cases for the treated range of temperatures, this value does not exceed a few Pa, which is significantly below both experimental and theoretical evaluation of adhesion of the material of the cometary nucleus (Blum et al. 2006). We can conclude that quasi-stationary sublimation of admixtures cannot destroy the surface material and peel the crystallized ice off.

A most promising idea was recently submitted by Gonzalez et al. (2008). The authors found that the crystallisation may have non-monotonic character for some range of model parameters. As a result, the crystallization front evolves discontinuously and crystallisation waves may be responsible for the outbursts. In the model by Gonzales et al. (2008) only pure water ice was considered, and questions concerning the activity of minor species ($CO$, $CO_2$) as well as surface evolution due to general erosion were not investigated. We plan to develop this idea further and consistently include into consideration: i) a model of radiative transfer in porous media, ii) a kinetic model of gas transport through porous media, iii) a model of crystallisation of amorphous water ice, and finally iv) a model of transient erosion of surface layers.

Summing up, we conclude that i) the total upper surface of the cometary nucleus should be peeled off continually in order to satisfy the observational results for the long-term high activity of the comets considered in this paper, ii) the most popular idea of cristallization of amorphous water ice cannot be responsible (at least in the current form) to explain the observed long-time cometary activity at the large heliocentric distance.

**Acknowledgments**

The observations, made at the 6-m telescope BTA, were performed due to the support of the Schedule Committee for Large Telescopes (Russian Federation). Grateful acknowledgment is made to Dr. Klaus Jockers for interisland discuss, constructive criticism. We thank A. Milani and CARA project members, E. Balducci, E. Bryssink, M. Nicolini, R. Ligustri, G. Sostero, for providing AFρ measurements. Collaboration work was done by the support of the grant DAAD.

**References**

Afanasiev, V.L., Moiseev, A.V. 2005. The SCORPIO universal focal reducer of the 6-m telescope. Astr. Lett. 31, 194–204.

A'Hearn, M., Schleicher, D. 1984. Comet Bowell 1980b. Astron. J. 89, 579–591.

Arvesen, J.C., Griffin, R.N., Jr., and Pearson, B. D. Jr. 1969. Determination of extraterrestrial solar spectral irradiance from a research aircraft. Appl. Opt. 8, 2215–2230.

Bar-Nun, A., Herman, G., Laufer, D., Rappaport, M. L. 1985. Trapping and release of gases by water ice and implications for icy bodies. Icarus. 63, 317-332.

Bar-Nun, A. and Laufer, D., 2003. First experimental studies of large samples of gas-laden amorphous "cometary" ices. Icarus. 161, 157-163.

Bauer, J.M., Fernández, Y., R., Meech, K. J. 2003. An Optical Survey of the Active Centaur C/NEAT (2001 T4). Astronomical Society of the Pacific. 115, 810, 981-989.

Biver, N., Rauer H., Despois D., Moreno R., Paubert G., Bockelee-Morvan D., Colom P., Crovisier J., Gerard E., and Jorda L. 1996. Substantial outgassing of CO from Comet Hale-Bopp at large heliocentric distance. Nature. 380, 137–139.

Blum, J., Schräpler, R., Davidsson, B., Trigo-Rodriguez, J. 2006. The Physics of Protoplanetesimal Dust Agglomerates I. Mechanical Properties and Relations to Primitive Bodies in the Solar System, Astrophys. J., 652, 1768-1781.

Bockelée-Morvan, D., Crovisier, J., Mumma, M. J., Weaver, H. A. 2004. The composition of cometary volatiles. Comets II. Univ. of Arizona Press, Tucson, AZ, pp. 391-423.

Busarev, V.V. 1999. The spectrophotometry of atmosphereless celestial bodies of the Solar system. Astron. Vestnik 33, 140–150 (Rus.).

Cochran, A.L, Barker, E.S., Cochran, W.D., 1980. Spectrophotometric observations of 29P/Schwassmann-Wachmann 1 during outburst. Astron. J. 85, 474–477.

Cochran, A.L, Cochran, W.D., Barker, E.S. 1982. Spectrophotometry of comet Schwassmann–Wachmann 1. II – Its color and $CO^+$ emission. Astrophys. J. 254, 816–822.

Cruikshank, D.P., Brown, R.H., 1983. The nucleus of comet 29P/Schwassmann–Wachmann 1. Icarus. 56, 377–380.

Enzian, A., Cabot, H., Klinger, J. 1997. A 2 1/2 D thermodynamic model of cometary nuclei. I. Application to the activity of comet 29P/Schwassmann-Wachmann 1. Astron. Astrophys. 319, 995-1006.

Epifani, M., Palumbo P., Capria M. T., Cremonese G., Fulle M., and Colangeli L. 2009. The distant activity of the Long Period Comets C/2003 O1 (LINEAR) and C/2004 K1 (Catalina). Astron. Astrophys. 502, 355-365.

Fernandez, J. A., Jockers, K. 1983. Nature and origin of comets. Progress in Physics. 46, 6, 665-772.

Fulle, M. 1992. Dust from short-period Comet P/Schwassmann-Wachmann 1 and replenishment of the interplanetary dust cloud. Nature. 359, 6390, 42-44.

Fulle, M., Cremonese, G., Böhm, C. 1998. The Preperihelion Dust Environment of C/1995 O1 Hale-Bopp from 13 to 4 AU. Astron. J., 116, 3, 1470-1477.

Green, D.W.E. 2003. Comet C/2002 VQ_94 (LINEAR). IAU Circ. 8194.

Greenberg, J. M. 1982. What are comets made of - A model based on interstellar dust. Comets, pp. 131-163.

González, M., Gutiérrez, P. J., Lara, L. M., Rodrigo, R. 2008. Evolution of the crystallization front in cometary models. Effect of the net energy released during crystallization. Astron. Astrophys. 486, 331-340.

Gronkowski, P., Smela, J. 1998. The cometary outbursts at large heliocentric distances . Astron. Astrophys. 338, 761-766.

Gronkowski, P., 2005. The source of energy of the comet 29P/Schwassmann-Wachmann 1 outburst activity: the test of the summary. Monthly Notices. 360, 3, 1153-1161.

Gronkowski, P. 2007. The search for a cometary outbursts mechanism: a comparison of various theories. Astronomische Nachrichten. 328, 126.

Davidsson, B. J. R., Skorov, Yu. V. 2004. A practical tool for simulating the presence of gas comae in thermophysical modeling of cometary nuclei. Icarus. 168, 1, 163-185.

Herget, P., 1961. The orbit of comet Schwassmann–Wachmann 1. Astron. J. 66, 266–271.

Houpis, H. L. F., Mendis, D. A., 1981. Dust emission from comets at large heliocentric distances. I - The case of comet Bowell /1980b. Moon and Planets. 25, 397-412.

Howell, S.B., Koehn, B., Bowell, E., Hoffman, M. 1996. Detection and measurement of poorly sampled point sources imaged with 2-D arrays. Astron. J. 112, 1302–1311.

Hughes, D. W. 1992. Cometary nuclei. Astron. Now. 6, 7, 41 – 43.

Jewitt, D., Meech, K. J. 1986. Cometary grain scattering versus wavelength, or 'What color is comet dust?' Astrophys. J. 310, 937-952.

Jewitt, D. 1990. The persistent coma of Comet P/Schwassmann-Wachmann 1 Astrophys. J. 351, 277-286.

Jewitt, D., Matthews, H. 1999. Particulate Mass Loss from Comet Hale-Bopp. The Astron. J. 117, 2, 1056-1062.

Jewitt, D., 2005. A first look at the Damocloids. Astron. J. 129, 530–538.

Jockers, K., Bonev, T., Ivanova, V., Rauer, H. 1992. First images of a possible CO(+)-tail of Comet P/Schwassmann-Wachmann 1 observed against the dust coma background. Astron. Astrophys. 260, 1-2, 455-464.

Jockers, K., Bonev. T. 1997. $H_2O^+$, $CO^+$, and dust in Comet P/Swift-Tuttle. Astron. Astrophys. 319, 617–629.

Jockers, K. 1997. Observations Of Scattered Light From Cometary Dust And Their Interpretation. Earth, Moon, and Planets. 79, 1/3, 221-245.

Ivanova A. V., Korsun P. P., Afanasiev V. L. 2009. Photometric investigations of distant comets C/2002 VQ94 (LINEAR) and 29P/Schwassmann-Wachmann-1. Sol. Syst. Res. 43, 5, 453-462.

Kartasheva, T.A., Chunakova, N.M., 1978. Spectral atmospheric transparency at the Special Astrophysical Observatory of the USSR Academy of Science in 1974 to 1976. Astrof. Issled. Izv. Spets. Astr. obs.10, 44 -51.(Rus.).

Korsun, P.P., Chorny G.F. 2003. Dust tail of the distant Comet C/1999 J2 (Skiff). Astron. Astrophys. 410, 1029–1037.

Korsun, P.P., Ivanova, O.V., Afanasiev, V.L., 2006. Cometary activity of distant object C/2002 VQ94 (LINEAR). Astron. Astrophys. 459, 977-980.

Korsun, P.P., Ivanova, O.V., Afanasiev, V.L., 2008. C/2002 VQ94 (LINEAR) and 29P/Schwassmann-Wachmann 1 - $CO^++N_2^+$ rich comets . Icarus 198, 2, 465-471.

Korsun P.P., Kulyk I.V., Ivanova O.V., Afanasiev V.L., Kugel F., Rinner C., Ivashchenko Yu.N. Dust tail of the active distant comet C/2003 WT42 (LINEAR) studied with photometric and spectroscopic observations. Icarus. (Submitted).

Kouchi, A. and Sirono S. 2001. Crystallization heat of impure amorphous $H_2O$ ice. Geophys. Res. Lett. 28, 827-830.

Landolt, A. U. 1992. Broadband UBVRI photometry of the Baldwin-Stone Southern Hemisphere spectrophotometric standards. Astron. J. 104, 1, 372-376.

Larson, S. 1980. $CO^+$ in comet Schwassmann–Wachmann 1 near minimum brightness. Astrophys. J. 238L, L47, L48.

Larson, S., Sekanina, Z., 1984. Coma morphology and dust–emission pattern of periodic Comet Halley. I – High–resolution images taken at Mount Wilson in 1910. Astron. J. 89, 571–578.

Larson, S., Brownlee, J., Hergenrother, C., Spahr, T. 1998. The Catalina sky survey for NEOs. Bull. Am. Astron. Soc. 30, 1037.

Li, A., Greenberg, J. M. 1997. A unified model of interstellar dust. Astron. Astrophys. 323, 566-584.

Lowry, S.C., Fitzsimmons, A. 2005. William Herschel telescope observations of distant comets. Mon. Not. R. Astron. Soc. 358, 641–650.

Marsden, B.G., 2002. MPEC 2002-V71.

Marsden, B.G. 2004. Comet C/2003 WT42 (LINEAR), MPEC 2004-D22.

Meech, K.J, Belton, M.J.S., Mueller, B.E.A., Dicksion, M.W., Li, H.R., 1993. Nucleus properties of 29P/Schwassmann–Wachmann 1. Astron. J. 106, 1222–1236.

Meech, K. J., Pittichová, J., Bar-Nun, A., Notesco, G., Laufer, D., Hainaut, O. R., Lowry, S. C., Yeomans, D. K., Pitts, M. 2009. Activity of comets at large heliocentric distances pre-perihelion. Icarus. 201, 2, 719-739.

Moreno, F. 2009. The Dust Environment of Comet 29P/Schwassmann-Wachmann 1 From Dust Tail Modeling of 2004 Near-Perihelion Observations. Astrophys. J. Supp. 183, 1, 33-45.

Mukai, T. 1986. Analysis of a dirty water-ice model for cometary dust. Astron. Astrophys. 164, 397–407.

Mumma M., DiSanti M., Dello Russo N., Magee-Sauer K., Gibb E., Novak R. 2003. Remote infrared observations of parent volatiles in comets: A window on the early solar system. Advances in Space Research.31, 12, 2563-2575.

Neckel, H. Labs, D. 1984. The solar radiation between 3300 and 12500 Å. Sol. Phys. 90, 205–258.

Oke, J.B. 1990. Faint spectrophotometric standard stars. Astron. J. 99, 1621-1631.

Parker, J. W. 2003. Comet Observations [695 Kitt Peak]. MPC. 49390, 9.

Patashnick, H. 1974. Energy source for comet outbursts, Nature. 250, 313-314.

Pravdo, S.H., Rabinowitz, D.L., Helin, E.F., Lawrence, K.J., Bambery, R.J., Clark, C.C. Groom, S.L., Levin, S., Lorr, J., Shaklan, S.B., Kervin, P., Africano, J.A., Sydney, P., Soohoo, V. 1999. The Near-Earth Asteroid Tracking (NEAT) program: An automated system for telescope control, wide-field imaging, and object detection. Astron. J. 117, 1616–1633.

Prialnik, D. and Bar-Nun A., 1990. Gas release in comet nuclei. Astrophys. J. 363, 274-282.

Prialnik, D., Bar-Nun, A. 1992. Crystallization of amorphous ice as the cause of Comet P/Halley's outburst at 14 AU. Astron. Astrophys. 258, 2, L9-L12.

Prialnik, D. 1992. Crystallization, sublimation, and gas release in the interior of a porous comet nucleus Astrophys. J., 388, 196-202.

Prialnik, D. 1997. Modelling Gas and Dust Release from Comet Hale-Bopp. Earth, Moon, and Planets. 77, 3, 223-230.

Prialnik, D., Sarid, G., Rosenberg, E. D., Merk, R. 2008. Thermal and Chemical Evolution of Comet Nuclei and Kuiper Belt Objects. Space Science Reviews. 138, 1-4, 147-164.

Rauer, H., Arpigny, C., Boehnhardt, H., Colas, F., Crovisier, J., Jorda, L., Küppers, M., Manfroid, J., Rembor, K., Thomas, N. 1997. Optical observations of comet Hale-Bopp (C/1995 O1) at large heliocentric distances before perihelion. Science. 275, 5308, 1909 – 1912.

Rettig, T.W., Tegler, S.C., Pasto, D.J., Mumma, M. J. 1992. Comet outbursts and polymers of HCN. Astrophys. J. 398, 1, 293-298.

Roemer, E. 1958. An outburst of comet Schwassmann–Wachmann 1. PASP 70, pp. 272–278.

Roemer, E. 1962. Activity in Comets at Large Heliocentric Distance. Publications of the Astronomical Society of the Pacific. 74, 440, 351.

de Sanctis, M. C., Capria, M. T., Coradini, A. 2002. Thermal evolution models of objects at large heliocentric distance. Proceedings of Asteroids, Comets, Meteors - ACM 2002. 92-9092-810-7, 39 – 42.

Sekanina, Z. 1975. A study of the icy tails of the distant comets. Icarus. 25, 218–238.

Sekanina, Z. 1982. Comet Bowell (1980b); an active-looking dormant object? Astron. J. 87, 161–169.

Sekanina, Z. 1996. Activity of Comet Hale-Bopp (1995 O1) beyond 6 AU from the Sun. Astronomy and Astrophysics.314, 957-965.

Sandford, S. A. and Allamandola, L. J. 1990. The volume- and surface-binding energies of ice systems containing CO, CO2, and H2. Icarus. 87, 188-192.

Skorov, Yu. V., Rickman, H. 1995. A kinetic model of gas flow in a porous cometary mantle. Planetary and Space Science. 43, 1587-1594.

Skorov, Yu. V., Kömle, N. I., Keller, H. U., Kargl, G., Markiewicz, W. J. 2001. A Model of Heat and Mass Transfer in a Porous Cometary Nucleus Based on a Kinetic Treatment of Mass Flow. Icarus. 153, 1, 180-196.

Smoluchowski, R. 1981. Amorphous ice and the behavior of cometary nuclei, Astrophys. J. 244, L31-L34.

Stansberry, J.A., Van Cleve, J., Reach, W.T., Cruikshank, D.P., Emery, J.P., Fernandez, Y.R., Meadows, V.S., Su, K.Y.L., Misselt, K., Rieke, G.H., Young, E.T., Werner, M.W., Engelbracht, C.W., Gordon, K.D., Hines, D.C., Kelly, D.M., Morrison, J.E., Muzerolle, J. 2004. Spitzer observations of the dust coma and nucleus of 29P/Schwassmann–Wachmann 1. ApJSS. 154, 463–468.

Stokes, G.H., Evans, J.B., Viggh, H.E.M., Shelly, F.C., Pearce, E.C. 2000. Lincoln Near- Earth Asteroid Program (LINEAR). Icarus. 148, 21–28.

Szabó, Gy. M., Kiss, L. L., Sárneczky, K. 2008. Cometary Activity at 25.7 AU: Hale-Bopp 11 Years after Perihelion. Astrophys. J. 677, 2, L121-L124.

Trigo-Rodríguez, J. M., García-Melendo, E., Davidsson, B. J. R., Sánchez, A., Rodríguez, D., Lacruz, J., de Los Reyes, J. A., Pastor, S. 2008. Outburst activity in comets. I. Continuous monitoring of comet 29P/Schwassmann-Wachmann 1. Astron. Astrophys. 485, 2, 599-606.

Whipple, F. L., 1972. Cometary Nuclei – Models. Comets: Scientific Data and Missions. 4.

Whipple, F.L., 1980. Rotation and outbursts of comet 29P/Schwassmann–Wachmann 1. Astron. J. 85, 305–31.

**Figure captions**

Figure 1. Schematic view of the Earth's, Jupiter's, Saturn's orbits, and a part of the comet orbits, located above the ecliptic plane. Positions of the comets for the observational dates, the moment of the perihelion passage, and the discovery date are marked.

Figure 2. Cropped images of comets 29P/Schwassmann-Wachmann 1 (left), C/2003 WT42 (LINEAR) (center) and C/2002 VQ94 (LINEAR) (right) observed through broad band filters. North, East, and the sunward directions are indicated.

Figure 3. Log-log dependence of the $Af\rho$ parameters on the aperture radius for comets 29P/Schwassmann-Wachmann 1, C/2003 WT42 (LINEAR) and C/2002 VQ94 (LINEAR) projected at the comets distances.

Figure 4a. Dependence of the dust production rate Q parameters on the aperture radius for comet 29P/Schwassmann-Wachmann 1 projected at the comet distance.

Figure 4b. Dependence of the dust production rate Q parameters on the aperture radius for comets C/2003 WT42 (LINEAR) and C/2002 VQ94 (LINEAR) projected at the comets distances.

**1) Table**

Table 1
Journal of observations

| Data, UT | r, AU | Δ, AU | Exp., s | $Z^{\alpha}$ | $P^{b}_{sun}$ | Fil | Comet |
|---|---|---|---|---|---|---|---|
| Dec., 15.919, 2006 | 5.865 | 4.947 | 200 | 31 | 102.4 | V | SW1 |
| Dec., 15.922, 2006 | 5.865 | 4.947 | 200 | 32 | 102.4 | V | SW1 |
| Dec., 15.924, 2006 | 5.865 | 4.947 | 100 | 33 | 102.4 | R | SW1 |
| Dec., 15.926, 2006 | 5.865 | 4.947 | 100 | 34 | 102.4 | R | SW1 |
| Dec., 15.999, 2006 | 5.865 | 4.947 | 200 | 35 | 102.4 | V | SW1 |
| Dec., 16.002, 2006 | 5.865 | 4.947 | 200 | 53 | 102.4 | V | SW1 |
| Dec., 16.005, 2006 | 5.865 | 4.947 | 100 | 54 | 102.4 | R | SW1 |
| Dec., 16.006, 2006 | 5.865 | 4.947 | 100 | 55 | 102.4 | R | SW1 |
| Mar., 29.817, 2009 | 6.117 | 5.789 | 120 | 36 | 100.1 | V | SW1 |
| Mar., 29.843, 2009 | 6.118 | 5.789 | 120 | 37 | 100.1 | V | SW1 |
| Mar., 30.352, 2009 | 6.118 | 5.805 | 120 | 37 | 100.1 | V | SW1 |
| Mar., 30.389, 2009 | 6.118 | 5.805 | 120 | 38 | 100.1 | V | SW1 |
| Mar., 30.427, 2009 | 6.118 | 5.805 | 60 | 43 | 100.1 | R | SW1 |
| Mar., 30.465, 2009 | 6.118 | 5.805 | 60 | 44 | 100.1 | R | SW1 |
| Mar., 30.647, 2009 | 6.118 | 5.805 | 60 | 44 | 100.1 | R | SW1 |
| Mar., 30.465, 2009 | 6.118 | 5.805 | 60 | 45 | 100.1 | R | SW1 |
| Dec., 13.415, 2006 | 7.084 | 7.332 | 180 | 36 | 320.8 | V | VQ94 |
| Dec., 13.475, 2006 | 7.084 | 7.332 | 180 | 35 | 320.8 | V | VQ94 |
| Dec., 13.537, 2006 | 7.084 | 7.332 | 180 | 34 | 320.8 | V | VQ94 |
| Dec., 13.615, 2006 | 7.084 | 7.332 | 120 | 34 | 320.7 | R | VQ94 |
| Dec., 13.661, 2006 | 7.084 | 7.332 | 120 | 33 | 320.7 | R | VQ94 |
| Dec., 13.701, 2006 | 7.084 | 7.332 | 120 | 33 | 320.7 | R | VQ94 |
| Dec., 16.758, 2006 | 5.519 | 5.598 | 30 | 32 | 300.1 | R | WT42 |
| Dec., 16.793, 2006 | 5.519 | 5.598 | 30 | 32 | 300.1 | R | WT42 |
| Dec., 16.814, 2006 | 5.519 | 5.598 | 60 | 32 | 300.1 | R | WT42 |
| Dec., 16.850, 2006 | 5.519 | 5.598 | 60 | 31 | 300.1 | R | WT42 |
| Dec., 16.886, 2006 | 5.519 | 5.598 | 60 | 31 | 300.1 | R | WT42 |

$^{\alpha}$ Zenith distance in degrees
$^{b}$ Position angle of the extended radius vector in degrees.

**Table 2**

Integral mass loss for the comets

| Comet | dusty | | Icy | |
|---|---|---|---|---|
| | rich | standard | Rich | Standard |
| SW1 | Total mass loss, $10^{11}$ kg | | | |
| | 2.47 | | 4.94 | |
| | Thickness of depleted layer, cm | | | |
| | 5.9 | 20.5 | 7.0 | 21.5 |

| WT42 | Total mass loss, $10^{11}$ kg | | | |
|---|---|---|---|---|
| | 0.99 | | 1.99 | |
| | Thickness of depleted layer, cm | | | |
| | 2.4 | 8.2 | 2.8 | 8.7 |
| VQ94 | Total mass loss, $10^{11}$ kg | | | |
| | 0.11 | | 0.21 | |
| | Thickness of depleted layer, cm | | | |
| | 0.3 | 0.9 | 0.3 | 0.9 |

**2) Figure**

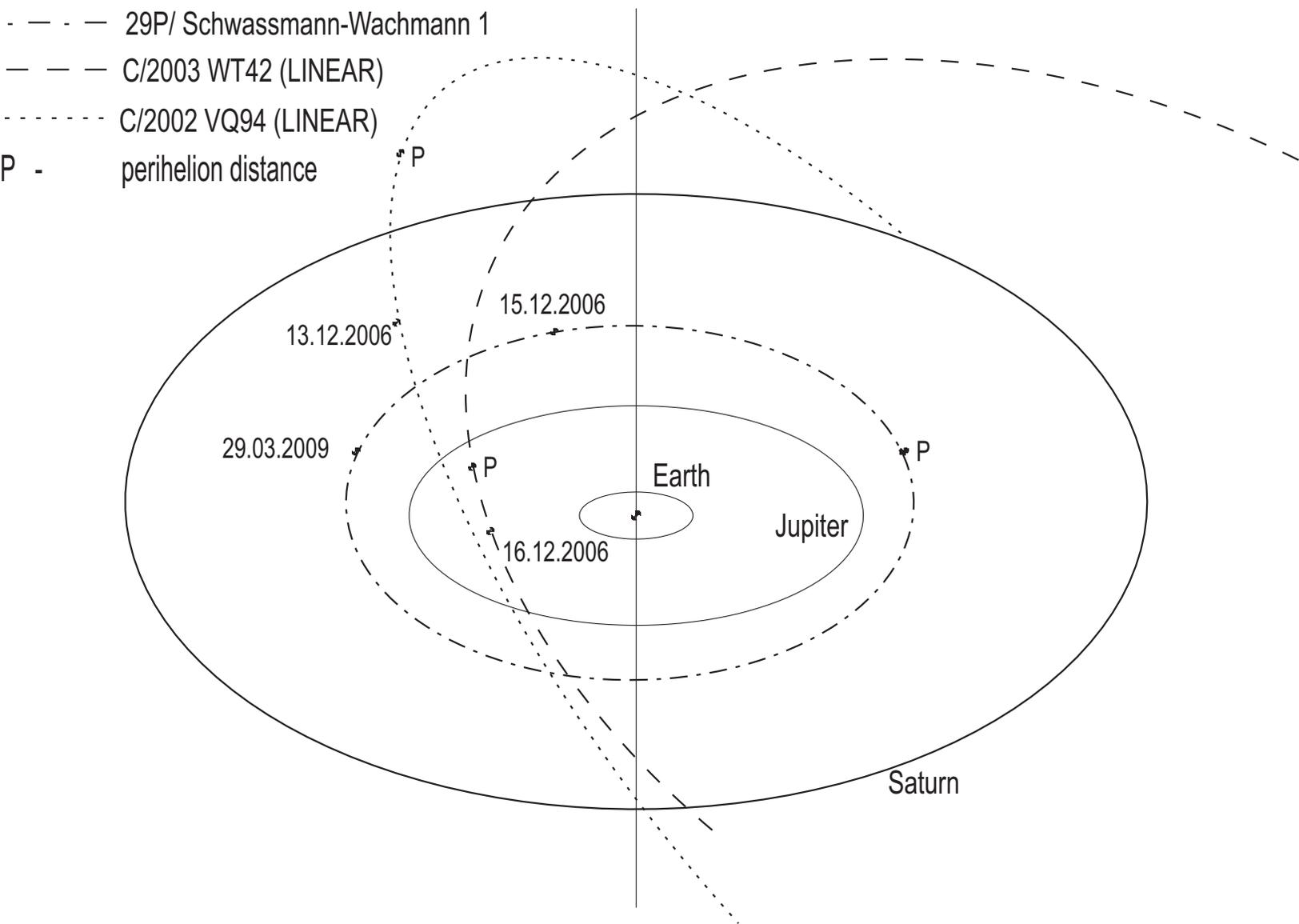

**2) Figure**

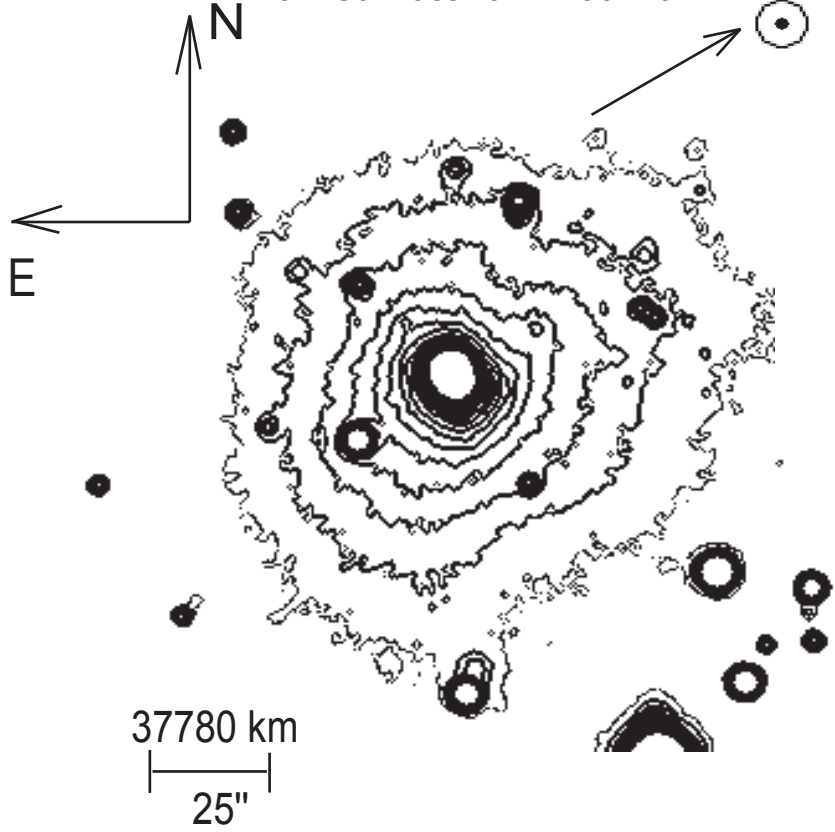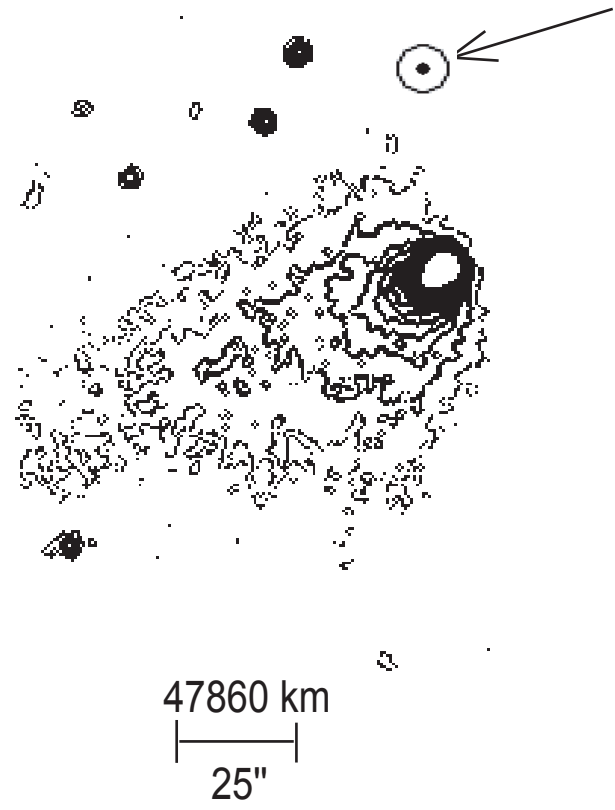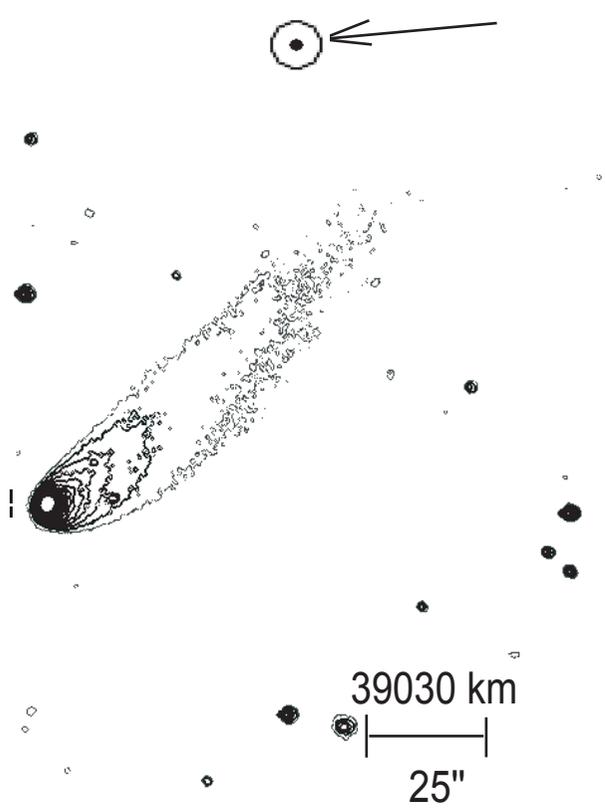

**2) Figure**

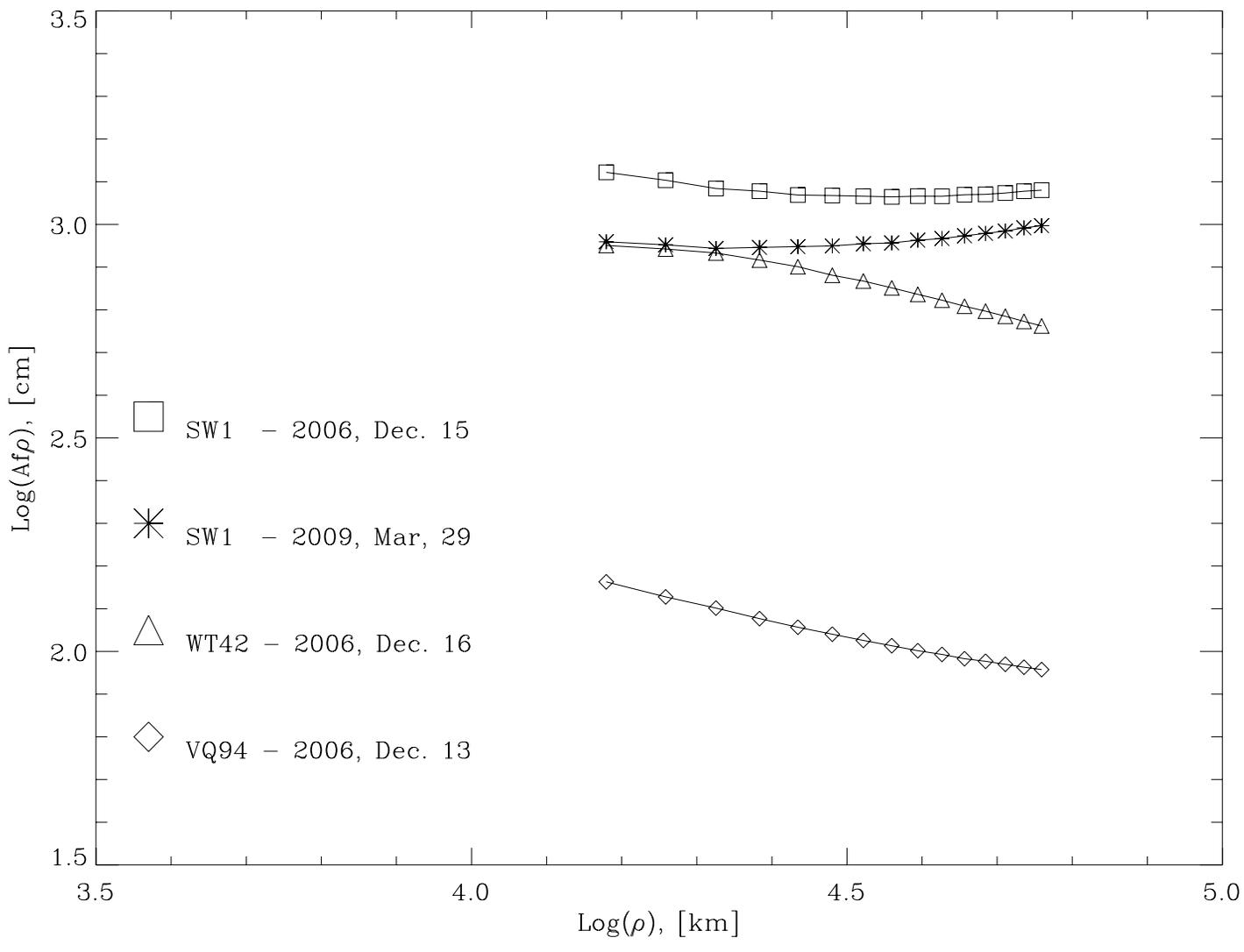

**2) Figure**

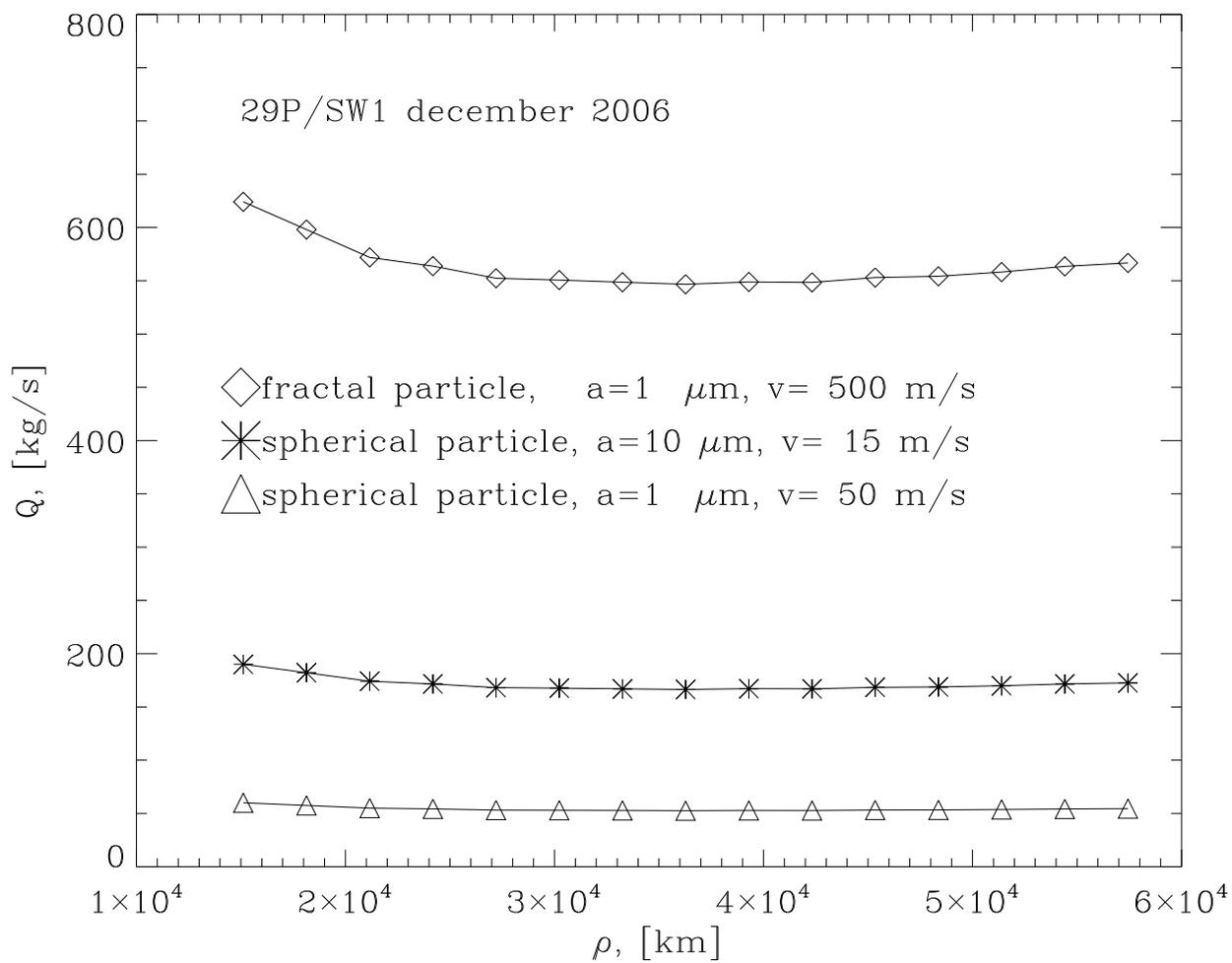

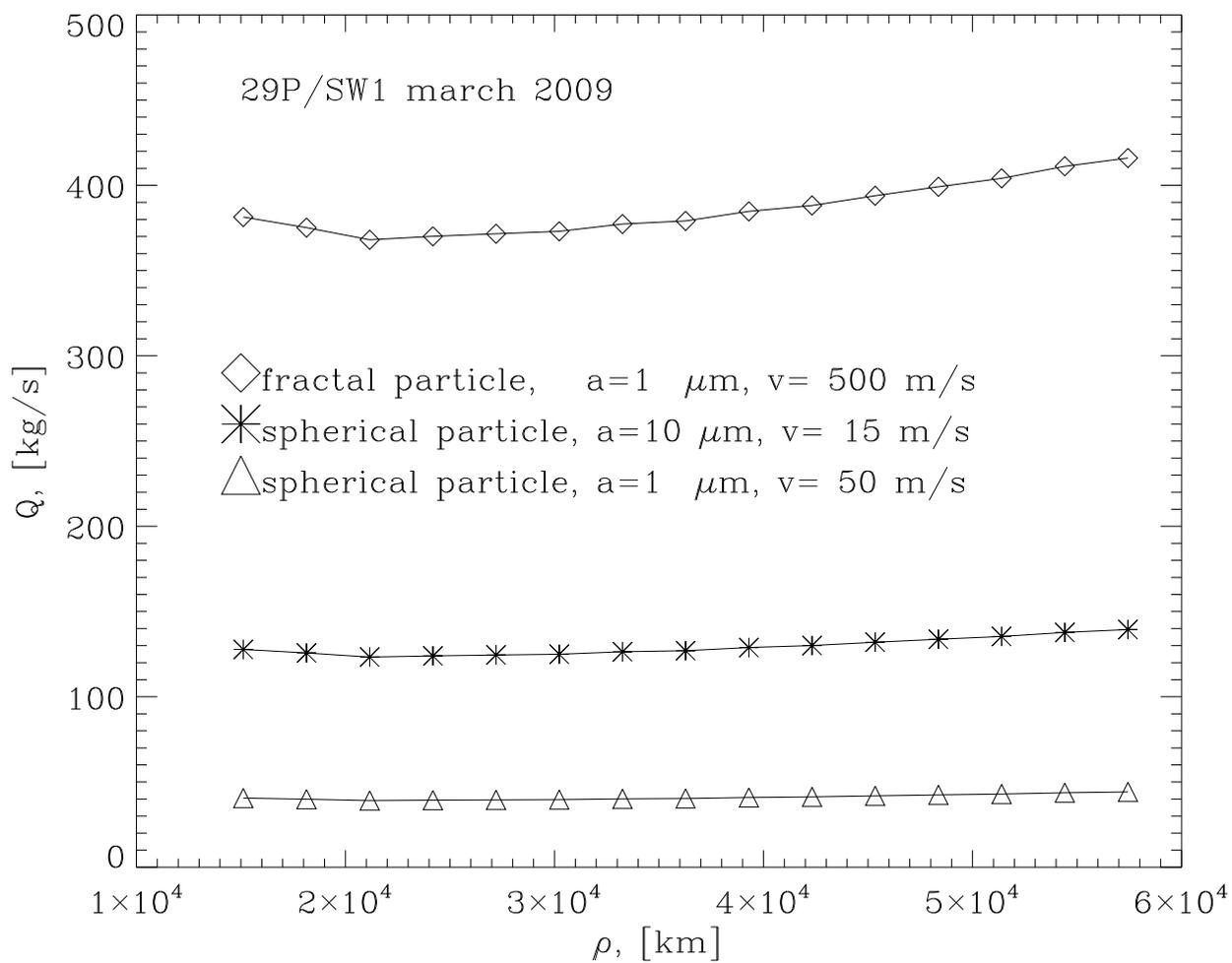

**2) Figure**

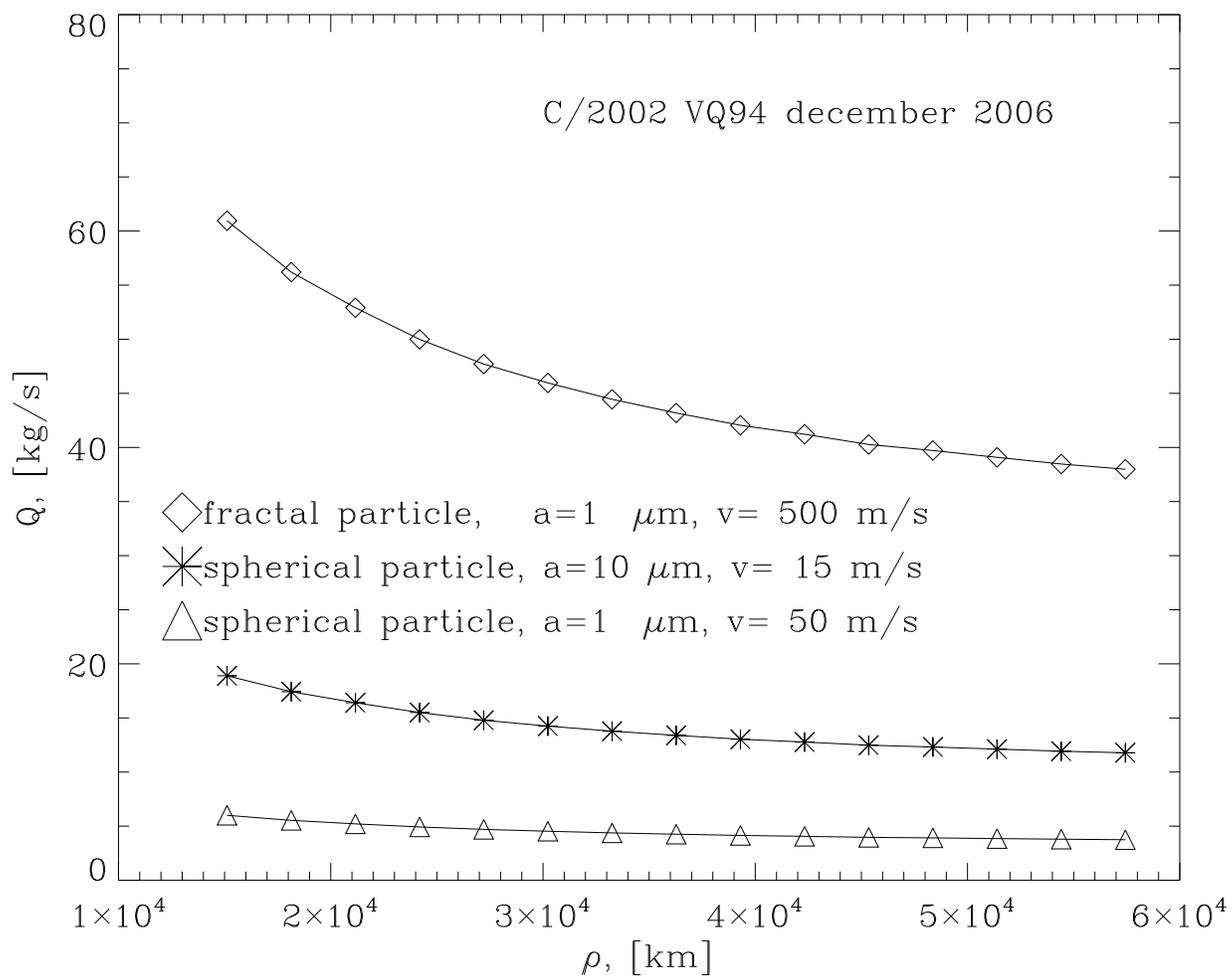

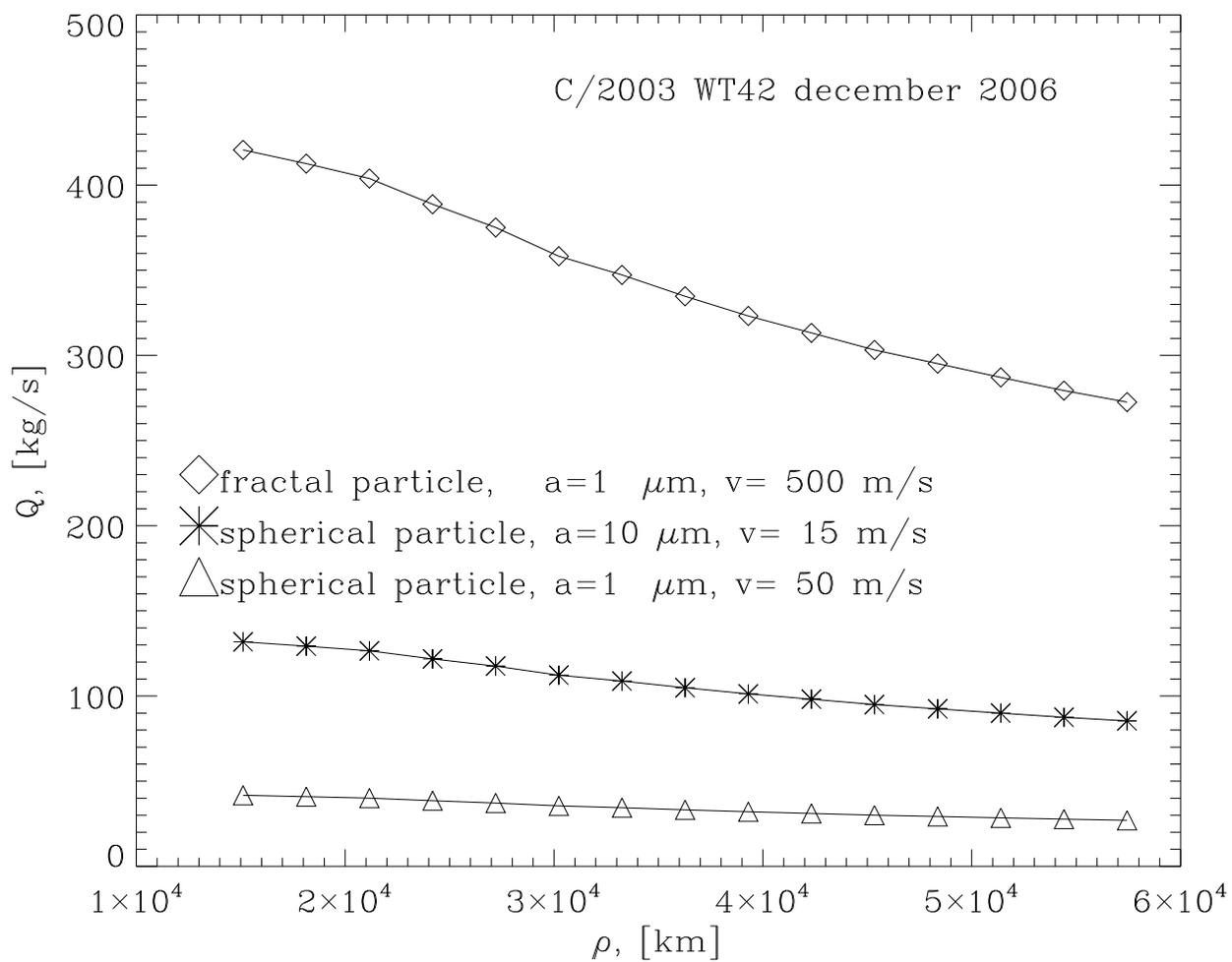